\documentclass[showpacs,preprintnumbers]{revtex4}
\usepackage{amsmath,amssymb,pstricks,wasysym,stmaryrd,enumerate,epsfig}

\begin{document}
\title{ Spectral density in resonance region and analytic confinement}
\author{Alex C. Kalloniatis\footnote{akalloni@physics.adelaide.edu.au} }

\affiliation{
Special Research Centre for the Subatomic Structure of Matter,
University of Adelaide,
South Australia 5005, Australia}

\author{Sergei N. Nedelko \footnote{nedelko@thsun1.jinr.ru}}
\affiliation{Institute of
Theoretical Physics~III of
Erlangen-Nuremberg University, Erlangen, Germany,
and
Bogoliubov Laboratory of Theoretical Physics, JINR,
141980 Dubna, Russia}

\author{Lorenz von Smekal \footnote{smekal@theorie3.physik.uni-erlangen.de}}
\affiliation{Institute of Theoretical Physics~III of
Erlangen-Nuremberg University, Erlangen, Germany}

\date{\today}
\preprint{ADP-04-14/T596}

\begin{abstract}
We study the role of finite widths of resonances 
in a nonlocal version of the Wick-Cutkosky model.
The spectrum of bound states is
known analytically in this model and forms linear Regge tragectories.
We compute the widths of resonances, calculate the spectral density in 
an extension of the Breit-Wigner
{\it ansatz} and discuss a mechanism for
the damping of unphysical exponential growth of observables
at high energy due to finite widths of resonances.
\end{abstract}
\maketitle

\section{Introduction}

Behaviour in field theories with nonlocal interactions in the 
time-like momentum region is the central concern of this paper.
Nonlocal quantum field theory has a long history.
Initially it was invented as an
attempt at solving the problem of UV divergences. However after the
development of renormalisation techniques for local QFT the study of 
nonlocal interactions became
a field of academic interest.
Nevertheless the challenge of constructing self-consistent and
mathematically rigorous quantum field theories with nonlocal interactions
and their phenomenological application
have been attracting permanent attention.
As a result of this long activity of several decades
the general principles of quantisation of nonlocal field theories
have been formulated.
Various issues related to quantisation, unitarity of the $S$-matrix,
realisation of causality, the validity of Froissart-type bounds at high energy
and the physical interpretation of nonlocal fields  can be found for instance
in~
\cite{Meiman,Fainberg,Efimov,Efi01,Moffat,Woodard,Leutw80,Efiv,Feyn,Blokh,Smekal}.
For the present paper the essential result is that
for a theory with typical growth of amplitudes in the
complex momentum plane and $\Lambda$ the scale of nonlocality
the Froissart type bound on
the total cross-section~\cite{Efi01},
\begin{eqnarray}
\label{bound}
\sigma_{\rm tot}(s) \leq C s^{\kappa-1} \log^2 s ,
\end{eqnarray}
\begin{eqnarray*}
{\cal A} \sim e^{ (p^2/\Lambda^2)^\kappa}, \ \ \kappa\ge 1/2
\end{eqnarray*}
can be derived assuming
\begin{itemize}
\item unitarity of the $S$-matrix,
\item analyticity of amplitudes on the Martin-Lehmann ellipse.
\end{itemize}
These two requirements are fulfilled in local as well as in nonlocal QFT with
the form factors being entire analytical functions of momenta.
Despite this upper bound on the total-cross section, practically useful
procedures for calculating observables at high energy in nonlocal QFT have
been not formulated.
Thus, since amplitudes are exponentially growing within perturbation theory,
the common wisdom is that nonlocal theories necessarily
display pathological behaviour at high energy,
which is usually considered as a fundamental drawback of
nonlocal QFT models~\cite{Smekal, Joglekar}.
The aim of the present paper is to give an example of such a
practical calculation for the total
annihilation cross section in a confining nonlocal model
exhibiting a Regge spectrum of bound states.
In order to decode this statement of the problem we first
have to explain its context.

An essential disadvantage of nonlocal theories has always been seen
in the functional ambiguity in the choice of form factors
related to the nonlocal character of interactions.
Though requirement of unitarity restricts the form factors
to be entire analytical functions of momenta,
an ambiguity within this class of form factors is unavoidable 
as far as nonlocality is introduced as a fundamental property of the theory.
An unexpected source for
a nonlocal character of quantum fields has been pointed out
by Leutwyler~\cite{Leutw80} who has noticed that the propagator of
the massless
charged scalar field in the presence of background (anti-)self-dual
homogeneous gauge field takes the form
\begin{eqnarray}
\label{scal}
D(p^2)=\frac{1-e^{-p^2/\Lambda^2}}{p^2},
\end{eqnarray}
here $p$ is Euclidean momentum, and scale $\Lambda$ is 
related to the strength of the background gauge field.
The nonlocal form factor $(1-\exp(-p^2/\Lambda^2))$ appears here due to the
presence of strong nonperturbative gauge fields, and thus its appearance 
itself and particular form
can be related to the properties of the
physical vacuum of the system. For vanishing field strength 
$\Lambda\to0$, and the usual scalar propagator is recovered.
Such an observation was immediately recognised as potentially
interesting in application to the confinement problem in QCD~\cite{Leutw80}.
For physical interpretation it is important that the
nonperturbative background gauge field eliminates the pole in the
momentum space propagator of a charged field rendering it
an entire analytical function. Such an analytical property of the propagator 
can be regarded as
dynamical confinement of the charged field since it has no particle
interpretation~\cite{Leutw80,Finjord,Efiv}. It is also important that 
such well established perturbative
properties of QCD as asymptotic freedom at short distances are unaffected by
the nonlocality generated by such background gluon fields.
We have here an example of a theory defined by a local classical
Lagrangian but with nonlocal behaviour of the
quantised fields appearing as a consequence of a nonperturbative gauge 
field configuration.
The (anti-)self-dual homogeneous gluon field
itself can be seen as a highly idealised analytically tractable 
realisation of more complicated
and realistic nonperturbative gauge configurations responsible for confinement
(for instance, the domain model~\cite{NK2001,NK2004} is a specific attempt at 
realising a more detailed picture of the QCD vacuum in this spirit).

Turning to confinement specifically now, this phenomenon
is more than just the absence of the coloured
quark-gluon states in the physical spectrum of QCD, but
the mechanism by which colourless bound states are present in 
the spectrum of QCD.
One of the first connections between confinement and the hadron spectrum
was achieved by the string picture, which provided
a simple explanation for the pattern of Regge trajectories
from a mechanism of confinement.
However, the string explanation for Regge trajectories is not unique.
Another, purely field theoretical approach to bound state formation
is provided by the hadronisation procedure in the QCD functional integral
which is closely connected with Bethe-Salpeter approach to the
description of relativistic bound states
\cite{PRC87,PRD95}.  It has been shown
within the bosonisation procedure that nonlocality in 
quark propagators similar to
Eq.~(\ref{scal}) generated by the
homogeneous (anti-)self-dual field in QCD leads to asymptotically linear  
Regge trajectories
~\cite{PRD95}. In fact this approach turned out to be successful 
in application 
to wide variety of meson spectra -- light, heavy-light mesons and 
heavy quarkonia~\cite{PRD96}.
The specific chiral properties of quark eigenmodes are
similar to those in an instanton
background and are of crucial importance in the
description of light mesons in this context.

In its most clean and refined form the role of
nonlocality of the type Eq. (\ref{scal}) in generating a
linear Regge spectrum of bound states
has been exposed in \cite{EfGan} where the method of
\cite{PRD95} has been improved and applied to a scalar field
model.  Unlike genuine QCD with all its complications, 
this model allows for analytical solution of the
Bethe-Salpeter equation in the one-boson exchange approximation. 
The Lagrangian of the model has the same structure
as that of the well-known Wick-Cutkosky model \cite{WickCutkosky} 
\begin{equation}
\label{lagr}
{\cal L} = - \Phi^{\dagger} S^{-1}(-\partial^2) \Phi
-\frac{1}{2} \phi D^{-1}(-\partial^2) \phi
- g \Phi^{\dagger} \Phi \phi.
\end{equation}
The Wick-Cutkosky model corresponds to a standard 
real massless scalar field $\phi$
and massive charged scalar field $\Phi$. It was invented as a prototype 
theory for studying the relativistic bound state problem in 
quantum electrodynamics. It turned out that the choice of
Euclidean momentum space propagators of  
the form
\begin{equation}
\label{props}
S(p^2)
= D(p^2)
= {1\over{\Lambda^2}} e^{-p^2/\Lambda^2}
\end{equation}
leads to a Bethe-Salpeter equation which is analytically soluble in the 
one-boson exchange approximation
generating a Regge spectrum of relativistic bound states~\cite{EfGan}
with mass-squared
\begin{eqnarray}
M_{nl}^2=M_0^2 + (2n+l) \ln(2+\sqrt{3})^2\Lambda^2,
\end{eqnarray}
which is linear both in radial number $n$ and angular momentum $l$.
Thus the model given by Eqs.(\ref{lagr}) and (\ref{props}) can be seen as a 
soluble prototype of a confining theory, here 
with confined fundamental fields 
$\phi$ and $\Phi$ and a Regge spectrum
of relativistic bound states representing the physical particle spectrum.
It is remarkable that the effective action for the composite
fields describing the bound states can also be derived 
analytically within the bosonisation approach,
thus enabling further calculations of physical quantities such as 
form-factors, decay widths, and scattering cross-sections.
The more realistic choice
\begin{equation}
\label{props-real}
S(p^2)\sim D(p^2)\sim \frac{1- e^{-p^2/\Lambda^2}}{p^2},
\end{equation}
where nonlocality appears as an (exponentially small at short
distances) correction to the local massless propagators, 
allows a variational approximate solution of the corresponding 
Bethe-Salpeter equation
and displays an asymptotically linear Regge spectrum~\cite{EfGan}.

In this paper we consider the
total annihilation cross section in the model of
Eqs.(\ref{lagr}) and (\ref{props}) using the method suggested in
\cite{Shif}, where a similar problem has been studied
in application to the 't~Hooft model. The total cross section for
annihilation of a lepton-anti-lepton pair
into hadrons via a scalar ``photon'' was evaluated in \cite{Shif}
by means of  continuation  of  Breit-Wigner
formulae in the complex plane
using the fact that the physical spectrum of the 't~Hooft model
consists of a Regge spectrum of hadronic resonances with the finite widths.
In our case the bound states form a physical Regge spectrum of resonances
and the effective ``hadron'' action is known explicitly. 
The scattering problem can be formulated in a standard way for these
hadrons and a unitary $S$-matrix can be defined. The
total cross section is calculated by means of the optical theorem,
namely via the imaginary part
of the correlator of two scalar currents
\begin{eqnarray}
\label{polar}
\Pi(q)=\int d^4x e^{iqx}\langle0| j(x) j(0)|0 \rangle,  
\ \ j(x)=\Phi^\dagger(x) \Phi(x).
\end{eqnarray}
As in the t'Hooft model the crucial point is to incorporate
into the calculation of the RHS of Eq.(\ref{polar})
the finite widths of the resonances of the physical spectrum.
As explained below, the imaginary part of the correlator is calculated
in the approximation based on an extended Breit-Wigner formula. 
This approximation is valid if resonances are
narrow enough, which is satisfied in the model under
consideration for the seven lowest
resonances corresponding to the energy interval 
$s\lesssim 30 \div 40 \Lambda^2$.

It should be stressed that our task is not verification of the
asymptotic inequality Eq.(\ref{bound}).
At asymptotically high energy the number of resonances involved in
the calculation of the total cross-section grows rapidly,
their width becoming so large that they strongly overlap.
In this regime an approximation based on the
Breit-Wigner formula is not valid and methods in the spirit of 
the statistical bootstrap \cite{Hag} should be more
appropriate.
As already stated, we intend rather to give a particular example
demonstrating  a
computation of physical quantities at relatively
high energies within a nonlocal model with confinement of fundamental fields
and a  Regge spectrum of bound states.

The final result for our  estimate of the spectral density is given in the
upper plot of Fig.\ref{fig:sigma}. We conclude that in a
nonlocal theory with confined fundamental fields and
exponentially growing amplitudes,
taking into account the physical bound states
can  suppress
unphysical growth of observable quantities at higher energies.
The finite width of these resonances is of particular importance. 
 Exponential growth of the widths is a
particularly distinctive feature of the analytic confinement scenario.  
It leads to the convergence of the Breit-Wigner type
series for the spectral function. In a broader context, as indicated in  
a recent paper~\cite{BlB03}, exponentially growing widths
of hadronic resonances can be important for understanding the 
phenomenology related to the quark-gluon plasma: 
it eliminates the divergence of the thermodynamic functions above 
the Hagedorn temperature and is able
to successfully describe lattice QCD data for the energy density. 
We discuss this issue briefly in the last section.

%%%%%%%%
\begin{figure}[htb]
\vspace{17mm}
\includegraphics{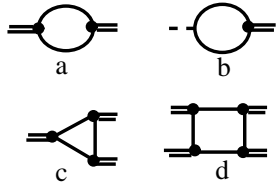}
\caption{Graphical representation of various terms in the effective action.}
\label{fig:diagr1}
\end{figure}
%%%%%

\section{The Effective Action for Composite Fields. }

Our starting point is the Euclidean functional integral
\begin{eqnarray}
\label{fint0}
{\cal Z}[I]={\cal N}\int {\cal D}\phi{\cal D}\Phi{\cal D}\Phi^\dagger
\exp\left\{\int d^4x \left [ {\cal L}(x) - \Phi^\dagger(x)\Phi(x) I(x)\right]  \right\},
\end{eqnarray}
where the Lagrangian ${\cal L}(x)$ is defined by 
Eqs.(\ref{lagr}) and (\ref{props}), and
the external current $I(x)$
\begin{eqnarray*}
I(x)=e^2\int d^4y G(x-y)\bar l(y) l(y)
\end{eqnarray*}
represents
interaction with the scalar lepton current $\bar l(y) l(y)$ by virtue 
of scalar particle (analogous to photon)
exchange reflected in the propagator $G(z)$.

The propagators $S(p^2)$
and $D(p^2)$ given in Eq.(\ref{props})
are entire functions such that  no
particles can be associated with the fields  $\Phi$
and  $\phi$. These fields represent fluctuations localised in space and time.
A typical space-time size of fluctuations is set by a confinement scale
$\Lambda$.
The physical particle spectrum of the system can be recovered by means of
the ``bosonisation'' procedure applied to the functional integral 
Eq.(\ref{fint0}).
For a detailed description of this procedure for this case we refer
the reader  to \cite{EfGan}.
Essentially, bosonisation requires solving the Bethe-Salpeter 
equation in the one-boson exchange approximation, which
can be done analytically due to the simple Gaussian form of propagators.
The main steps are as follows. Integrating out the field  $\phi$
one arrives at a quartic interaction of the scalar fields $\Phi$:
\begin{equation}
L_2[\Phi] = \frac{g^2}{2} \int d^4x_1 d^4 x_2
\Phi^{\dagger}(x_1)\Phi(x_1) D(x_1-x_2) \Phi^{\dagger}(x_2)\Phi(x_2).
\end{equation}
Introducing a complete orthonormal set of functions
$U_{\cal Q}$, corresponding to the radial quantum number, 
total momentum and magnetic numbers  ${\cal Q}=\{n,l,\mu\}$,
the four point interaction can be rewritten
as an infinite sum of products of a non-local currents
\begin{equation}
L_2= \frac{g^2}{2} \sum_{\cal Q} \int d^4x J_{\cal Q}(x) J_{\cal Q}(x)
\end{equation}
with
\begin{eqnarray}
J_{\cal Q}(x) &=& \Phi^{\dagger} (x) 
V_{\cal Q}( \stackrel{\leftrightarrow}{\partial})\Phi(x) \nonumber \\
V_{\cal Q}(\stackrel{\leftrightarrow}{\partial}) &=& 
\int d^4y \sqrt{D(y)} U_{\cal Q}(y) e^{\frac{y}{2}
\stackrel{\leftrightarrow}{\partial}
}
\end{eqnarray}
One introduces auxiliary fields $\Psi_{\cal Q}$
with a subsequent Gaussian integration over
the field $\Phi$. As a result the
original functional integral (\ref{fint0}) turns out to be identically
rewritten in terms of composite fields $\Psi_{\cal Q}$ 
representing collective excitations in the system
\begin{eqnarray}
\label{fint1}
Z &=&
\prod_{\cal Q} \int {\cal D}\Psi_{\cal Q} \exp \left\{
-\frac{\Lambda^2}{2}  \int d^4p \Psi_{\cal Q}(-p) (\delta_{\cal Q Q'} -
\alpha \Sigma_{\cal Q Q'}(p)) \Psi_{\cal Q'}(p)
+ W_I[g \Psi]  \right\}
\end{eqnarray}
where the dimensionless coupling constant $\alpha = g^2/ (4\pi \Lambda)^2$, and
\begin{equation}
W_I[g\Psi] = - {\rm {Tr}} [ \ln(1- g\Psi_{\cal Q} V_{\cal Q} S- I S) +
\frac{g^2}{2} \Psi_{\cal Q} V_{\cal Q} S \Psi_{\cal Q'} V_{\cal Q'} S ]
\end{equation}
describes interactions between composite fields from which
the quadratic term in the expansion of the logarithm is
subtracted and added to
the original quadratic part,
\begin{equation*}
\alpha\tilde\Sigma_{\cal Q Q'}(x-y) = \frac{g^2}{2\Lambda^2} V_{\cal Q}(\stackrel{\leftrightarrow}{\partial}_x) S(x-y)
V_{\cal Q'}(\stackrel{\leftrightarrow}{\partial}_y) S(y-x)
\end{equation*}
The nontrivial requirement is that the basis $U_{\cal Q}$ be chosen such that
the self-energy $\Sigma_{\cal Q Q'}(p)$
and hence the quadratic part of the effective action is diagonal in 
the quantum numbers $Q$,
\begin{equation}
\Sigma_{\cal Q Q'}(p) = E_{\cal Q}(-p^2) \delta_{\cal Q Q'}
\end{equation}
Bound state masses are then real
solutions to
\begin{equation}
1=\alpha E_{\cal Q}(M_{\cal Q}^2).
\end{equation}
Diagonalisation of the self-energy is equivalent to the solution of the
Bethe-Salpeter equation in the one-boson exchange approximation.
In the momentum representation propagators and vertices involved in
the effective action have the form~\cite{EfGan}

%%%%%%%%
\begin{figure}[htb]
\vspace{30mm}
\includegraphics{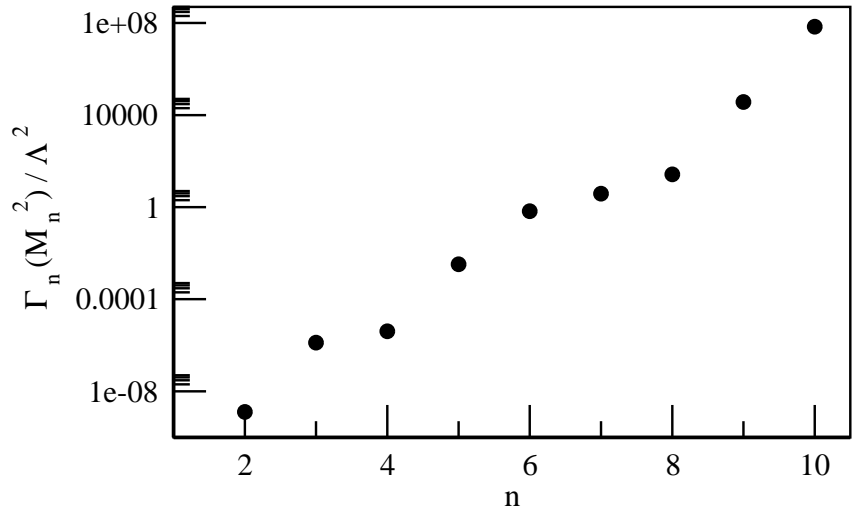}
\caption{The two-particle decay width
$\Gamma_{n}(M_n^2)$ as a function of radial quantum number. Deviation from
strictly exponential behaviour is due to the polynomial 
$P_n(s)$ in $\Gamma_{n}(s)\propto P_n(s)\exp({\rm const} \ s)$.}
\label{fig:width-n}
\end{figure}
%%%%%

\begin{eqnarray}
\label{vertex}
S(p)&=&\Lambda^{-2} e^{-p^2/\Lambda^2},
\nonumber\\
V^{nl}_{\mu_1\dots\mu_l}(K)&=&e^{i(q-p)x}\left[e^{ipx}V^{nl}_{\mu_1\dots\mu_l}\left(\stackrel{\leftrightarrow}{\partial}_x\right)e^{-iqx} \right],
\nonumber\\
&=&(-1)^{n+l}  \Lambda C_{n l}
T^l_{\mu_1, \dots, \mu_l} (K)
L^{l+1}_n(a K^2) e^{-b K^2}
\end{eqnarray}
with $K=(p+q)/2\Lambda$,  $a=2\sqrt{3}$,  $b=2/(1+\sqrt{3})=4/(2+a)$ and
\begin{eqnarray*}
C_{n l} = { {2^{n+2+2l} \sqrt{3}^{l/2+1}} \over
{(1+\sqrt{3})^{2n+l+2} }} \sqrt{ {{n! (l+1)}\over {(n+l+1)!}}}.
\end{eqnarray*}
Here the angular part of the vertex is given by 
$T^l_{\mu_1\dots\mu_l}$, the irreducible tensors of the Euclidean rotation 
group $O(4)$,
\begin{eqnarray}
&&\int_\Omega \frac{d\omega}{2\pi^2}T^{l}_{\mu_1\dots\mu_l}(n_y)T^{k}_{\nu_1\dots\nu_k}(n_y)
=\frac{1}{2^l(l+1)}\delta^{lk}\delta_{\mu_1\nu_1}\dots\delta_{\mu_l\nu_l}
\nonumber\\
&&T^{l}_{\mu_1\dots\mu\dots\nu\dots\mu_l}(n_y)=T^{l}_{\mu_1\dots\nu\dots\mu\dots\mu_l}(n_y),
\ T^{l}_{\mu\mu\dots\mu_l}(n_y)=0,
\nonumber\\
&&T^{l}_{\mu_1\dots\mu_l}(n_y)T^{l}_{\mu_1\dots\mu_l}(n_x)=\frac{1}{2^l}C_l^{(1)}(n_yn_x), \ \ n_x^2=n_y^2=1
\label{gegen}
\end{eqnarray}
with  $C_l^{(1)}$ being Gegenbauer polynomials.
The radial part corresponds to the Laguerre polynomials $L^{l+1}_n$,
with
\begin{eqnarray*}
\int_0^\infty du u^{l+1}e^{-u}L^{l+1}_n(u)L^{l+1}_{n'}(u)=\delta_{nn'}
\end{eqnarray*}
Calculation of the self-energy gives
\begin{eqnarray}
\label{Enl}
E_{nl}(-p^2) &=&  {e^{-{{p^2}\over {2\Lambda^2}}}\over {(2+\sqrt{3})^{2n+l+2}}},
\nonumber
\end{eqnarray}
and the square of the bound state masses read
\begin{equation}
M_{nl}^2=2 \Lambda^2 \left[\ln\frac{(2+\sqrt{3})^2}{\alpha} + (2n+l) \ln(2+\sqrt{3})\right]
\end{equation}
which manifests a linear Regge spectrum.
Finally the fields should be rescaled
\begin{eqnarray}
\label{hQ}
&&\Psi_{\cal Q}=\Psi_{\cal Q} g^{-1}\Lambda h_{\cal Q},
\nonumber\\
&& h^{-2}_{\cal Q}=
\frac{\Lambda^2}{(4\pi)^2}
\frac{d E_{\cal Q}(-p^2)}{dp^2}\Big\vert_{p^2=-M^2_{\cal Q}}
\end{eqnarray}
in order to ensure the correct residue of the propagator at the mass pole. 
 The coupling constant $h_{\cal Q}$ is defined as a dimensionless quantity.
Thus the final form of the functional integral for composite fields is
\begin{eqnarray}
\label{fint2}
Z &=&
\prod_{\cal Q} \int {\cal D}\Psi_{\cal Q} \exp \left\{
-\frac{\Lambda^4h^2_{\cal Q}}{2g^2}  \int d^4p \Psi_{\cal Q}(-p) (1 - \alpha E_{\cal Q}(-p^2)) \Psi_{\cal Q}(p)
+ W_I[h_{\cal Q} \Psi]  \right\}.
\end{eqnarray}
The original coupling constant $g$ enters only the
quadratic part of the effective action, while the remaining 
terms contain the effective coupling constant $h_{\cal Q}$.

The formal functional integral (\ref{fint2})
can be used to define a unitary nonlocal theory for the interacting fields 
$\Psi_{\cal Q}$ by defining the appropriate Gaussian measure for its 
computation. Namely, one rewrites $E_{\cal Q}(-p^2)$
as
\begin{eqnarray}
\label{measure}
&&1 - \alpha E_{\cal Q}(-p^2)=(M^2_{\cal Q}+p^2)\alpha E'_{\cal Q}(M^2_{\cal Q})
-\left[\alpha E_{\cal Q}(-p^2)  -1+ (M^2_{\cal Q}+p^2)\alpha E'_{\cal Q}(M^2_{\cal Q})\right],
\\
&&1 -\alpha E_{\cal Q}(M^2_{\cal Q})=0.
\nonumber
\end{eqnarray}
This decomposition exists since $E_{\cal Q}(-p^2)$ is an 
entire analytical function.
If one uses now the first term of the RHS of Eq.(\ref{measure}) 
to define the standard free field
Gaussian measure for computation of the functional integral then the 
theory of interacting
composite fields $\Psi_{\cal Q}$ so defined is unitary
since all nonlocal form factors in the action are entire analytical functions,
and the only singularities which appear in the $S$-matrix are related to the 
physical particles associated with fields $\Psi_{\cal Q}$.
In particular, this theory satisfies the conditions required for the
derivation of the high energy bound Eq.(\ref{bound}) with $\kappa=1$. 
Such a ``free field'' Gaussian measure is appropriate for computing
the processes
in which all composite particles are almost on-shell. In this case
$\left[\alpha E_{\cal Q}(-p^2)
-1+ (M^2_{\cal Q}+p^2)\alpha E'_{\cal Q}(M^2_{\cal Q})\right]=O((M^2_{\cal Q}+p^2)^2)$
is small and can be considered as a perturbation.

%%%%%%%%
\begin{figure}[htb]
\vspace{17mm}
\includegraphics{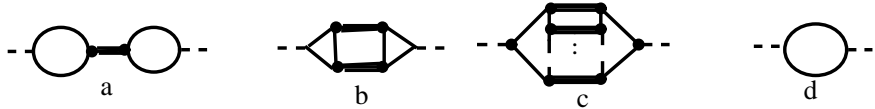}
\caption{Graphical representation of the polarisation function.}
\label{fig:diagr2}
\end{figure}
%%%%%

\section{Total cross section of  $\bar l l$ annihilation to ``hadrons''.}

According to the optical theorem the
total cross section for the inclusive process $\bar l l\to {\rm ``hadrons''}$
\begin{eqnarray*}
\sigma_{\rm tot}(s) =\frac{1}{2s} \sum_{\cal Q} (2\pi)^4 \delta^{(4)}(p_{\cal Q}-q)
|\langle {\cal Q}|T|\bar l l\rangle |^2
\end{eqnarray*}
can be expressed through the imaginary part of the 
amplitude for elastic forward scattering, which is proportional to the 
spectral density $\rho(s)$, namely the imaginary part of the 
correlator of two scalar currents
\begin{equation*}
\rho(s) =\Im \int d^4x e^{iqx}\langle0| j(x) j(0)|0 \rangle|_{q^2=s},  \ \ j(x)=\Phi^\dagger(x) \Phi(x).
\end{equation*}
In terms of the effective action in Eq.(\ref{fint1}) the spectral density
is given by the imaginary part of diagrams with one, two and generally
$n$ intermediate ``hadrons''
as shown in Fig.\ref{fig:diagr2}a-c. Since
the propagator of the constituent field $\Phi$  
(thin solid line) and the vertices $V_{\cal Q}$
(see Eq.(\ref{vertex})) are
entire analytical functions no contribution to
the imaginary part of diagrams in Fig.\ref{fig:diagr2} can come from the
constituent field loops.
In particular, the diagram in Fig.\ref{fig:diagr2}d has no
imaginary part and does not contribute to the spectral density.

%%%%%%%%
\begin{figure}[htb]
\vspace{30mm}
\includegraphics{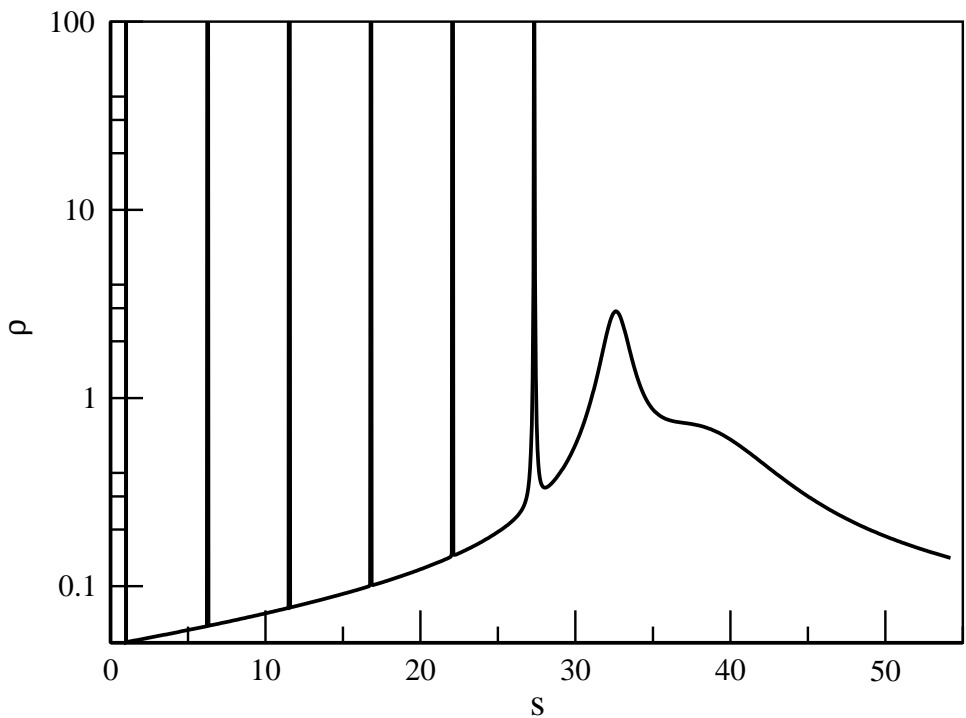}
\caption{Spectral density in the Breit-Wigner approximation as defined by
Eq.(\ref{densityBW}) }
\label{fig:rhoBW}
\end{figure}
%%%%%

In the Breit-Wigner approximation of the resonance propagator,
the contribution of diagram in Fig.\ref{fig:diagr2}a to the spectral density
takes the form
\begin{eqnarray}
\label{densityBW}
&&\rho_{BW}(s)=\sum_{n=0}^{\infty}\lim_{\varepsilon\to0}
\frac{h^2_{n0}g^2_{n0}(M^2_{n0})[\varepsilon+\Gamma_{n}(M^2_{n0})]}
{[(M^2_{n0}-s)]^2+[\varepsilon+\Gamma_{n}(M^2_{n0})]^2},
\end{eqnarray}
where $\Gamma_{n}(M^2_{n0})$ is the width of the resonance, which is
equal to zero for the lowest state.
It should be noted that the expansion analogous to that in
$1/N_{\rm c}$ in the 't Hooft model or $QCD$ is represented in the
model under consideration via an expansion in the
number of loops which include internal resonance propagators.
Further calculations will be based on this decomposition over ``hadronic'' 
loops.
At this point we take this expansion as a way of classifying the
various contributions, and will discuss its validity
in more detail shortly. 
As for the leading order contributions in $1/N_{\rm c}$ in the 't Hooft model,
the bound states in the nonlocal model are stable
in the zeroeth order of this expansion over ``hadronic'' loops.
In this approximation,
namely with decay widths neglected in Eq.(\ref{densityBW}),
the spectral density is simply an infinite equidistant sequence of 
delta functions. The lowest order contribution to the
widths of resonances in this loop expansion
are the two-particle decays. Numerically the widths for few lowest resonances
are shown in Fig.\ref{fig:width-n}. One can see that the width is 
exponentially growing with $n$, which ensures convergence of the 
series for the spectral density in Eq.(\ref{densityBW}) for any $s$
since $$h^2_{n0}g^2_{n0}\propto n$$ for $n\gg1$.
The exponential growth of the decay width is precisely due to nonlocality.
This should be compared with the behaviour  $\Gamma_n\propto\sqrt{n}$ and $g_{n}\to {\rm const}$ in the 't Hooft model 
or string based picture of confinement.
We comment in the discussion section on
whether available experimental data can distinguish between these two qualitatively different
dependencies.

Summing the convergent series in Eq.(\ref{densityBW}) gives the 
spectral density shown in Fig.\ref{fig:rhoBW}.
Qualitatively it has a reasonable form and
decreases at large $s$. The advantage of the Breit-Wigner approximation for
the resonance propagator is that it explicitly ensures the
correct analytical properties: only physical singularities in the 
complex $s$-plane are present in the Breit-Wigner propagator.

However, there are three important factors at large $s$ 
which are missed in the spectral density calculated through the
Breit-Wigner propagator with inclusion of two-particle
decay widths. First of all decays to three and more particles become 
important at large $s$. Accounting for them would broaden the resonances and 
lower the spectral density.
The second contribution comes from the diagrams with two or more intermediate
bound states (see Fig.\ref{fig:diagr2}b,c) which would increase the 
spectral density at large $s$.  These two factors indicate that 
Eq.(\ref{densityBW}) cannot give a truly reliable result
for asymptotically large $s$. At most it provides a hint at
the tendency of $\rho(s)$ to decrease at larger $s$.
The third factor is that the Breit-Wigner approximation is valid, 
strictly speaking, only in the vicinity
of a resonance pole and if the resonance is narrow enough.
This means for $s$ located away from
the resonance position the contribution of the $n$-th resonance to the
series Eq.(\ref{densityBW})  might not be reflected correctly.
In terms of the effective action discussed in the previous section
the Breit-Wigner approximation for the propagator corresponds
to the  ``free field'' Gaussian measure in the functional integral with
the addition of the on-shell widths of resonances. 
It is clear that corrections coming from the second term
in the RHS of Eq.(\ref{measure}) are not small for $p^2$ being off-shell. 
Certain resummations are necessary.

%%%%%%%%
\begin{figure}[htb]
\vspace{17mm}
\includegraphics{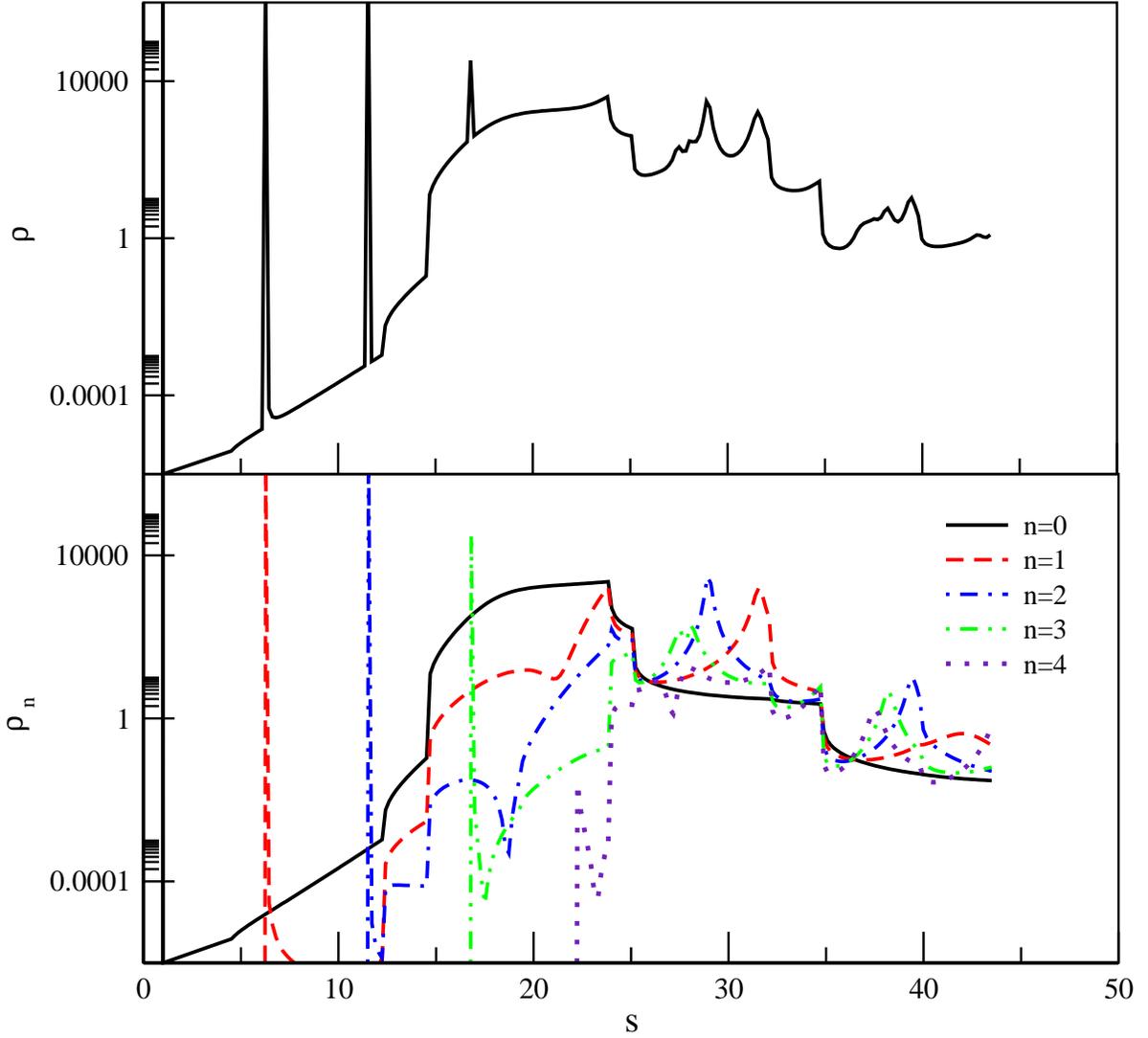}
\caption{Spectral density $\rho(s)$ as a function of the energy variable $s$ 
(upper plot)
and the contributions $\rho_n(s)$ of
few lowest radial numbers to the total spectral density (lower plot)}
\label{fig:sigma}
\end{figure}
%%%%%

The effective action at our disposal offers
an extension of the Breit-Wigner representation,
which can account for such resummations just mentioned
via inclusion of
the energy dependence of the self-energy $E_{n0}(s)$,
the transition amplitude $g_{n0}(s)$ and the
imaginary part -- decay width $\Gamma_{n0}(s)$.
This extension of the spectral density takes the form
\begin{eqnarray}
\label{density}
&&\rho(s)=\sum_{n=0}^{\infty}\rho_n(s), \
\nonumber\\
&& \rho_n(s)=\lim_{\varepsilon\to0}
\frac{|g_{n0}(s)|^2[\varepsilon+\tilde\Gamma_{n0}(s)]}{[1-\alpha E_{n0}(s)]^2+[\varepsilon+\tilde\Gamma_{n0}(s)]^2},
\nonumber\\
&&\tilde\Gamma_{n0}=\frac{16\pi^2\alpha}{h_{n0}\Lambda^2}\Gamma_{n0}
\end{eqnarray}
where the summation spans only radial excitations with $l=0$
because of conservation of orbital
momentum which is a good quantum number in this problem.
The transition amplitude $g_{n0}(s)$ is given by the diagram in
Fig.\ref{fig:diagr1}b.
Simple calculation  gives
\begin{equation}
\label{gns}
g_{n0}(s) =  (-1)^n \sqrt{n+1}E_{n0}(s), \
E_{n0}(s)=\frac{e^{s/2\Lambda^2}}{(2+\sqrt{3})^{2n+2}}.
\end{equation}
The quantity $\Gamma_{n0}(s)$ is the total width of two-particle decays of a
resonance with orbital momentum $l=0$,  radial number $n$ and
squared-energy  $s$.
Details of the calculation of $\Gamma_{n0}(s)$ are given in the appendix.

%%%%%%%%
\begin{figure}[htb]
\vspace{17mm}
\includegraphics{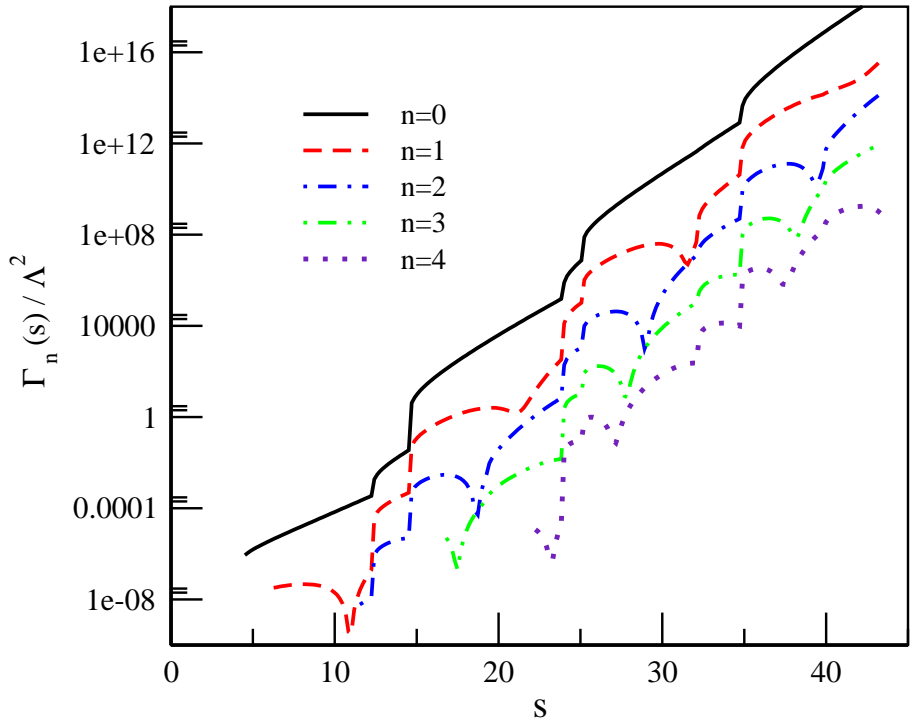}
\caption{Two-particle decay widths $\Gamma_{n}(s)$ as a function of 
energy variable $s$ for the few lowest resonances.}
\label{fig:width}
\end{figure}
%%%%%

Numerical results for $\Gamma_{n0}(s)$ are given in Fig.\ref{fig:width},
which shows its exponential growth with $s$, and
\begin{eqnarray*}
\Gamma_{n0}(s)\gg g_{n0}(s) \propto E_{n0}(s)
\end{eqnarray*}
for large $s$.  This inequality results in the behaviour of
the separate terms $\rho_n(s)$ shown in the lower plot in Fig.\ref{fig:sigma}.
Summing up these terms gives the spectral density shown in the upper plot in
Fig.\ref{fig:sigma}.
Qualitatively this spectral density reproduces the character of the 
simple Breit-Wigner approximation, Fig.\ref{fig:rhoBW}. 
The oscillations seen in $\rho_n(s)$ in Fig.\ref{fig:sigma} additional to
the resonance peaks are reflections of two features.
The first is the corresponding oscillations 
in $\Gamma_n(s)$ in  Fig.\ref{fig:width} which are due to the opening of 
new decay channels at certain values of $s$. 
The second, and more important, cause is the functional structure of 
the amplitude ${\cal A}_n(s)$ (see appendix):  essentially it is a product 
of a polynomial and exponent in $s$.
Again the spectral density shows several sharp peaks corresponding to 
the lowest narrow resonances,
then it has a maximum and a tendency to decrease at higher energy.
 It should be stressed that $\rho(s)$ is limited by the region of
not very large values of $s$ because of  the same as for $\rho_{\rm BW}(s)$  two reasons
(decays into more than two final states and diagrams with two or more
intermediate resonances). However unlike
the plain Breit-Wigner approximation the extended
formulae for the propagator should be more appropriate for the values of $s$
away from resonance position
although only in the vicinity of the real axis in the complex $s$-plane.
The reason preventing it from being
a good approximation in the whole complex plane is hidden in the property of
entire functions which have infinitely many zeroes in the complex plane.
In addition to the desirable physical
pole corresponding to the resonance, the extended propagator has
artificial unphysical poles for complex values of $s$.
As soon as these poles are sufficiently far from the real axis and do not 
noticeably affect the form of $\rho(s)$ the
approximation Eq.~(\ref{density})
can be considered appropriate. 
In the absence of the
term with $\Gamma_n$ in Eq.~(\ref{density}) 
and using Eq.~(\ref{Enl}), the unphysical singularities would appear at
$s=M_{nl}^2+4i\pi k\Lambda^2$ for any
nonzero integer $k$. The addition of $\Gamma_n(s)$ shifts all poles to
the complex plane, and a cursory numerical check reveals  
that the artificial singularities
produced by the resummation are located  sufficiently far
from the real axis.

\section{Discussion}

In summary, we have presented the spectral density in the
Breit-Wigner approximation
for a nonlocal model with confined constituent fields but
a Regge spectrum of bound states.
The two-particle decay widths have been computed and incorporated into the
evaluation of the spectral density.
The main observations are: firstly, due to rapid growth of resonance widths 
with growing radial number the formal sum
over resonances (which represents the spectral density)
is convergent; secondly, apart from the expected delta-function like peaks 
corresponding to the lowest narrow resonances the spectral density 
in this approximation
decays at larger values of the energy variable
$s$ as shown in Fig.\ref{fig:rhoBW}.

The Breit-Wigner approximation roughly reproduces the
behaviour of the resonance propagator
as a function of $s$ away from the resonance position.
In order to improve this a resummation 
of the series over the effective ``bound state-constituent field'' 
coupling constants $h_{\cal Q}$
are necessary, what we have called an
 ``extended'' Breit-Wigner approximation.
 The spectral density computed
by means of such a resummation shows qualitatively the same behaviour as the
Breit-Wigner spectral function -- there are several peaks corresponding to the
lowest narrow resonances and a tendency to decay at higher energies.

As already discussed,
the resummation introduces artificial
unphysical singularities.
However, in the particular case under consideration
these unphysical singularities
are located sufficiently far from the real $s$-axis. Insofar as we
consider the resonance propagator in the vicinity of the real axis, the
approximation based on resummation of
higher orders in $h_{\cal Q}$ is quite reliable.
It should be noted that, in the context of nonlocal quantum field theory, 
the problem of unphysical singularities appearing in the complex $s$-plane
is akin to the Landau pole problem in local
quantum field theory: a partial resummation of the perturbation series
is generally hard to achieve without introducing unphysical singularities,
which applies both to local and nonlocal models. 
There are two aspects to this issue. The practical way to manage
is to look for summation prescriptions which are appropriate in the 
restricted region of the complex momentum plane relevant to a given problem;
this is what we have pursued in the present work. 
A more fundamental approach to the problem would require formulation 
of the general principles and methods of summation which would
avoid the appearance of unphysical singularities. 

As mentioned, the asymptotic dependence of two particle decay widths on 
the bound state quantum number in the nonlocal model
$\Gamma_{n}\propto\exp\{{\rm const} \ n\}$ is drastically different from the
dependence in the t'Hooft model $\Gamma_{n}\propto\sqrt{n}$,
and semi-classical string based estimations of \cite{Casher} where 
$\Gamma(M)\propto M$.
The computation in the 't Hooft model is similar to ours in that
decays of excitations to lower states at the same trajectory are
considered, while the computation of~\cite{Casher}
applies to all possible decays.
It is interesting to take a look at experimental data on decay widths
in the hope to distinguish between these qualitatively different situations. 
Unfortunately the available data seem to be inconclusive in this respect. 
The problem is that for the purpose
of extracting asymptotic dependence of the width on radial or orbital number
one should consider only decay modes of a given excited state to  
states lower on the same Regge trajectory.
In the real multiflavour world, decays of excited states are dominated 
by the modes into final states which include
ground state mesons from Regge trajectories of other flavour octet states.
A typical example is the family of orbital excitations of $K^*$,  
$$((K^*(892),K^*_2(1430),K^*_3(1780),K^*_4(2045))$$
with the total widths (using \cite{pdg}
and data available through the 2003 partial update)
$$
\Gamma_{K^*}=50\pm.9 {\rm MeV}, \ \Gamma_{K_2^*}=98.5\pm 2.9 {\rm MeV}, \Gamma_{K_3^*}=159\pm.21 {\rm MeV}, \Gamma_{K_4^*}=198\pm 30 {\rm MeV},$$
looking as well fitting linear in $l$ behaviour of the width
(though $\Gamma \propto e^{M}$ expected in the domain model 
\cite{NK2001,NK2004} would also fit but be
implausible with the limited range of data).
However the problem is that decay modes which contribute
to the widths of these resonances all include $K$, $\pi$, $\rho$, $\eta$
mesons which obviously do not belong to the trajectory of $K^*$.
So, this approximately linear dependence in $l$ does not
contain the information which we are interested in.
It can be strongly influenced by chiral dynamics, and its description requires
a model with both chiral symmetry and confinement
implemented simultaneously (for instance the model of \cite{PRD96} has
a chance to provide such
a suitable framework for calculating partial widths known from experiment and 
enabling thus a detailed analysis). 
The clearest case would be the Regge family of radial excitations of a
pseudoscalar meson like $\pi, \pi(1300), \pi(1800)$,
however the widths and especially the partial widths for different 
decay modes are not known for these mesons with the required accuracy.
These two examples are quite typical, and we conclude that it is 
not yet possible to distinguish between the two
asymptotic behaviours on the basis of experimental data. 
It should be stressed that the nonlocal model,
despite the exponential asymptotics for the width,
indicates very small partial widths of the few lowest resonances 
for decays into lower states on the
same trajectory due to the polynomial pre-exponent, and can considerably 
deviate from strictly exponential form  
as can be seen from Fig.~\ref{fig:width}. 
These partial modes would anyway be screened by other modes in the multiflavour
real world.  A further signature of exponential behaviour for
larger $(n,l)$ states  would be that observable trajectories cannot be
"long", since resonances with radial or orbital numbers bigger than some 
critical $n_c$ ($n_c=6$ in Fig.~\ref{fig:width}) or $l_c$ are too wide 
to be observable.
Such a sharp cut-off of the trajectories would be typical for 
exponential dependence, but not for linear or square root behaviour.
The property that the known Regge trajectories, for example, for
unflavoured mesons cut-off at about $2$ GeV may be an indication of
this feature.

The crucial role of the finite widths in modulating the behaviour of
the cross-section for growing energy can be understood in terms of the
energy being dissipated into the creation of unstable particles, which
decay into more stable particles and so on. In this respect, the
mechanism is close in spirit to the Hagedorn mechanism for a maximal
temperature in hadron-hadron collisions \cite{Hag} where one has no 
recourse to the optical theorem and thus the machinery of the
statistical bootstrap is necessary {\it a priori}. In our case, this framework 
would be useful once the narrow width approximation has broken
down.
 In this context we mention that exponentially growing decay
widths can impact on the deconfinement phase transition
in particular the persistence of hadronic properties beyond
the critical temperature. For example in \cite{BlB03},
an ansatz involving
$\exp(m/T_H)$ for the width contribution to the spectral function of a 
hadron gas at temperatures greater than the Hagedorn temperature,
$T_H$, reproduces the characteristic slow approach to the Stefan-Boltzmann
ideal quark-gluon gas result seen in lattice calculations
\cite{KRT03}. Interestingly, the Hagedorn temperature scale,
traditionally argued to be determined by the mass of the lightest
hadron, is commensurate with the confinement scale which figures
prominently in analytical confinement and the exponentially
growing decay widths discussed in this paper:
$\Lambda\sim T_H\sim 200 {\rm {MeV}}$. The possibility
that the confinement scale is more significant in
fixing the Hagedorn temperature, especially in the chiral limit,
has been argued recently \cite{BFS04}. Further study of the consequences 
of analytic confinement to high temperature
behaviour is appropriate.

Finally, we have not addressed
the relation between our study to the idea and formalism of 
``quark-hadron duality''. With the purely Gaussian propagators
treated in this article, such an investigation would be pointless.
However this problem becomes well-formulated if one considers the more
realistic propagators of the form Eq. (\ref{scal}).
This propagator has the same short distance behaviour as
the local one and nonlocal corrections are exponentially small in the deep
Euclidean region. With such propagators one can address the
issue of quark-hadron duality
within a model with dynamical confinement of constituent fields,
in a spirit close to that of \cite{Shif}.

\appendix
\section{Two-particle decay width.}
The width for two-particle decay
for a state with the time-like four momentum squared $p^2=s$
decaying to particles with masses $m_1$ and $m_2$
is given by the standard equation
\begin{equation}
\Gamma^{(2)}(s,m_1,m_2) = {1\over {2s}} {1\over {(2\pi)^{2}}}
\int \prod_{i=1,2} {{d^3p_i}\over{2E_i}} \delta^{(4)}(p- p_1-p_2)
  |A(p,p_1,p_2)|^2  . 
\end{equation}
The modulus of the amplitude is independent of the momenta. Thus
for the width for the decay of the state with angular momentum $l=0$ and   
radial number $n$  into the 
final states with masses $m_1$  and $m_2$ and
quantum numbers $(n_1,l_1)$ and $(n_2,l_2)$ respectively, we get 
\begin{eqnarray*}
&&\Gamma_{n,n_1l_1,n_2l_2}^{(2)} = {1\over {(2\pi)^{2}}} {R_2(s,m_1,m_2)\over {2s}}  T_{n,n_1l_1,n_2l_2}(s,m_1^2,m_2^2),
\nonumber\\
&&T_{n,n_1l_1,n_2l_2}=|A_{n,n_1l_1,n_2l_2}(p,p_1,p_2)|^2,
\end{eqnarray*}
where $R_2(s,m_1,m_2)$ is the two-particle phase space
\begin{eqnarray*}
&&R_2(s,m_1,m_2)= {{\pi}\over s} \lambda^{\frac{1}{2}} (s,m_1^2,m_2^2)
\\
&&\lambda (s,m_1^2,m_2^2)=(s-(m_1+m_2)^2)(s-(m_1-m_2)^2) .
\end{eqnarray*}
The total width is then
\begin{eqnarray*}
\Gamma_n^{(2)}(s) =
\sum_{n_1l_1,n_2l_1} \Gamma_{n,n_1l_1,n_2l_2}^{(2)}(s,m_1^2,m_2^2).
\end{eqnarray*}
The amplitude $ A_{n;n_1l_1,n_2l_2}$ which should be computed
is given by the diagram Fig.\ref{fig:diagr1}c. 
For arbitrary $(l_1,l_2)$ and polarisations, the corresponding loop integral
takes the form (all momenta below represent dimensionless ratios,
e.g. $p=p/\Lambda$)
\begin{eqnarray*}
A_{\mu_1\dots\mu_{l_1},\nu_1\dots\nu_{l_2}}^{n;n_1l_1,n_2l_2}=\Lambda h_{n0}h_{n_1l}h_{n_2l}
\int {{d^4 q}\over{(2\pi)^4}}
V^{n0}(q) S(q+p/2) V_{\mu_1\dots\mu_{l_1}}^{n_1 l_1}(q+p/2-k_1/2) S(q+p/2-k_1)
\\
\times V_{\nu_1\dots\nu_{l_2}}^{n_2 l_2}(q-k_1/2) S(q-p/2) .
\end{eqnarray*}
Since the orbital momentum of the incoming state is zero,
the final states are required to be on-shell, and the angular
momentum is a conserved quantum number in this problem, we need to extract
from this amplitude the contribution of final states with 
total angular momentum equal to zero.
Thus we need the  term with the angular momentum $l=0$ in the
decomposition of the amplitude 
$A_{\mu_1\dots\mu_{l_1},\nu_1\dots\nu_{l_2}}^{n;n_1l_1,n_2l_2}$
over the irreducible tensors $T^{l}$ with $l=0,\dots, l_1+l_2$.
The required contribution is proportional to the trace of
$A_{\mu_1\dots\mu_{l_1},\nu_1\dots\nu_{l_2}}^{n;n_1l_1,n_2l_2}$ for $l_2=l_1$
\begin{eqnarray*}
{\cal A}_{n;n_1,n_2;l_1}=\frac{\Lambda^{-1}}{l_1!}
\sum_{\mu_1\dots\mu_{l_1}}
A_{\mu_1\dots\mu_{l_1},\mu_1\dots\mu_{l_1}}^{n;n_1l_1,n_2l_1},
\end{eqnarray*}
where ${\cal A}$ is defined as a dimensionless quantity.
Taking into account Eqs.(\ref{vertex}) and (\ref{gegen})
we arrive at
\begin{eqnarray}
\label{ampl}
\nonumber \\
{\cal A}_{n;n_1,n_2;l}&=& \frac{(-1)^{n+n_1+n_2}}{l!2^l} h_{n0}h_{n_1l}h_{n_2l}C_{n0} C_{n_1 l} C_{n_2 l}\int \frac{d^4q}{(2\pi)^4}
C_l^{(1)}[(q+(p-k_1)/2)(q-k_1/2)]
\nonumber\\
&\times&L^1_n[aq^2]
L^{l+1}_{n_1}[a (q+(p-k_1)/2)^2]
L^{l+1}_{n_2}[a(q-k_1/2)^2] C_l^{(1)}[(q+(p-k_1)/2)(q-k_1/2)]
\nonumber \\
&\times& \exp\left\{
-b q^2 - b(q+(p-k_1)/2)^2 - b(q-k_1/2)^2
-(q+p/2)^2 - (q+p/2-k_1)^2 - (q-p/2)^2  \right\},
\end{eqnarray}
where $C_l^{(1)}$ are Gegenbauer polynomials.
The Gaussian integral over the loop momentum $q$
can be performed analytically.
The most convenient way to do this is to use
a generating function for Laguerre polynomials
\begin{eqnarray*}
\sum_{n=0}^{\infty} L^{\beta}_n(x) z^n = {1\over{(1-z)^{\beta+1}}}
e^{\frac{xz}{z-1}}.
\end{eqnarray*}
Both sides of Eq.(\ref{ampl}) are multiplied 
by $z^n z_1^{n_1} z_2^{n_2}$ and summed over
$n,n_1,n_2$ and the result of integration is expanded in a power series
in each of $z, \ z_1$ and $z_2$. The coefficients of expansion  
gives the amplitudes ${\cal A}_{n;n_1,n_2;l}$. 
The Gegenbauer polynomials are substituted in explicit form.
This procedure can be easily implemented in {\it Mathematica} or {\it Maple}.
Finally the amplitude ${\cal A}_{n;n_1,n_2;l}$ as a function of $s$ 
is a product of a polynomial in $s$ and an exponential of $s$. 
The degree of the polynomial depends on $n$, $n_1$, $n_2$ and $l$.

The total two-particle decay width of the resonance with 
four momentum squared $s$, zero orbital momentum
and radial number $n$ entering Eq.(\ref{density}) is then
\begin{eqnarray}
\label{2total}
\Gamma^{(2)}_n= {\Lambda^2\over {8\pi^2s}}
\sum_{l,n_1,n_2}R_2[s,M^2(n_1,l),M^2(n_2,l)]
|{\cal A}_{n;n_1,n_2;l}[s,M^2(n_1,l),M^2(n_2,l)]|^2,
\end{eqnarray}
where all states with nonzero phase space are summed.

\section*{Acknowledgements}
This work was supported by the University of Adelaide under the
Small Research Grants scheme and partially by the grant RFBR~04-02-17370.
ACK is supported by a Research Fellowship of the Australian
Research Council.
S.N.N was supported by the DFG, contract SM70/1-1.
We thank Garii Efimov, Rajeev Bhalerao, Tony Thomas, Rod Crewther,
Johann Rafelski and Frieder Lenz  for clarifying discussions
and Ian Cloet for helpful assistance.  David Blaschke is thanked for attracting our attention to the work \cite{BlB03}.

\end{document}